\documentclass[%
 reprint,
superscriptaddress,
 amsmath,amssymb,
 aps,
]{revtex4-2}
\usepackage{graphicx}
\usepackage{dcolumn}
\usepackage{bm}
\usepackage{braket}

\begin{document}
\title{Closing of gaps, gap labeling and passage from molecular states to critical states in a 2D quasicrystal}

\author{Anuradha Jagannathan}
\affiliation{
Laboratoire de Physique des Solides, Universit\'{e} Paris-Saclay, 91405 Orsay, France
}
\date{February 2023}

\begin{abstract}
The single electron spectrum and wavefunctions in quasicrystals continue to be a fascinating problem, with few known exact solutions. We investigate the energy spectra and gap structures for a quasiperiodic tiling in two dimensions. Varying a continuous parameter, we follow the evolution of the band structure from discrete molecular or atomic states, to the multifractal states well-known from previous studies. We propose a gap labeling scheme for finite approximants which is different from, but converges to, one introduced by Kellendonk and Putnam for infinite tilings.
\end{abstract}

\maketitle
\section{Introduction}
Electronic properties of quasicrystals and their topological characteristics, have been the subject of many recent studies, most of these related to the 1D Fibonacci chain \cite{akkermans,zilberberg,rai}. Less is known about higher dimensional quasicrystals. In this paper we focus on a archetypal 2D system, the octagonal (or Ammann-Beenker) tiling \cite{grunbaum}. This paper considers spectral gaps and pseudogaps which have posed a conundrum, and explains their physical origins, their locations and topological indices. 

The standard hopping model on the octagonal tiling is defined by 
\begin{eqnarray}
\label{eq:ham1}
    H_1 &=& -t \sum_{\langle i,j\rangle}  (c^\dag_i c_j + h.c.)  
\end{eqnarray}
where $\langle i,j\rangle$ denotes pairs of sites linked by an edge. The hopping amplitude $t$ is uniform. The onsite energies also are constant, and have been set to zero. The complexity of the problem thus arises solely due to the variations in the local environments. In this model, the ground state is known to be multifractal \cite{mace} and all other eigenstates are believed to be multifractal as well (see the review in \cite{grimmreview}), with the exception of a set of $E=0$ confined states \cite{oktel}. The density of states in this model has many pseudogaps (sharp dips, where the density of states approaches zero) whose origin has been unclear until now.

The integrated density of states (IDOS) is defined by
  $  I(E) = \frac{1}{N} \sum_a \Theta (E-E_a)$,
where $E_a$ are the eigenvalues and $\Theta$ is the Heaviside function.  Kellendonk and Putnam \cite{kelput} showed, using the algebraic structure of the octagonal tiling, that the IDOS in any given gap is given by $I(\mathrm{p},\mathrm{q}) = (\mathrm{p} + \mathrm{q}(2+ \sqrt{2}))/8\lambda$, where  $\mathrm{p}$ and $\mathrm{q}$ are integers. This labeling involves an irrational number, $\lambda=1+\sqrt{2}$, also known as the ``silver mean". It is the limit of $\frac{P_{n+1}}{P_n}$, as $n \rightarrow \infty$, where the Pell numbers $P_n$ obey the recursion relation $P_n=2P_{n-1}+P_{n-2}$ with $P_0=0, P_1=1$. The result of Kellendonk and Putnam extended to a 2D system the celebrated gap labeling theorem for 1D quasicrystals \cite{bellissardgap}. 

We propose here an indexing scheme for gaps in periodic approximants of the octagonal tiling.
These structures, which approach the perfect quasicrystal as $n$ tends to infinity, can be generated by a select-and-project method \cite{baakegrimm,duneau} (see Appendix). One can also transform one approximant to the next one by means of an inflation. This is the process of decimating sites of a tiling according to a well-defined rule, and getting a tiling of the same type with edges $\lambda$ times bigger. The number of sites in the $n$th ($n>0$) approximant is 
$Q_{2n+1}$. The numbers $Q_n$ are given by the same recursion relation as the Pell numbers, that is, $Q_n=2Q_{n-1}+Q_{n-2}$ with initial conditions $Q_0=1, Q_1=1$. The two series of numbers are related by $Q_n=P_n+P_{n-1}$. 

We show that a labeling scheme for gaps in the finite approximants is given by
\begin{align}
\label{eq:label}
I_\mathrm{q} = [ \mathrm{q} ~\frac{Q_{2n-1}}{Q_{2n+1}}]
\end{align}
where $[x]$ denotes Mod$(x,1)$, for $0\leq \vert q\vert \leq \mathrm{Int}[Q_{2n+1}/2]$ where $\mathrm{Int}$ denotes the integer part. Note that a single integer $q$ suffices to label the gap. The spectrum and gaps and indexation are found by a perturbation expansion analogous to that for the 1D Fibonacci chain due to \cite{kkl,niunori}. Our numerical results confirm these predictions, and show that outside the regime of validity of perturbation theory results tend smoothly towards the limiting case of Eq.1. Importantly, our approach sheds light on previously unexplained features of models for magnetism, superconductivity, or impurity screening on tilings.


\subsubsection{Molecular model A}
\begin{figure} 
    \includegraphics[width=0.7\columnwidth]{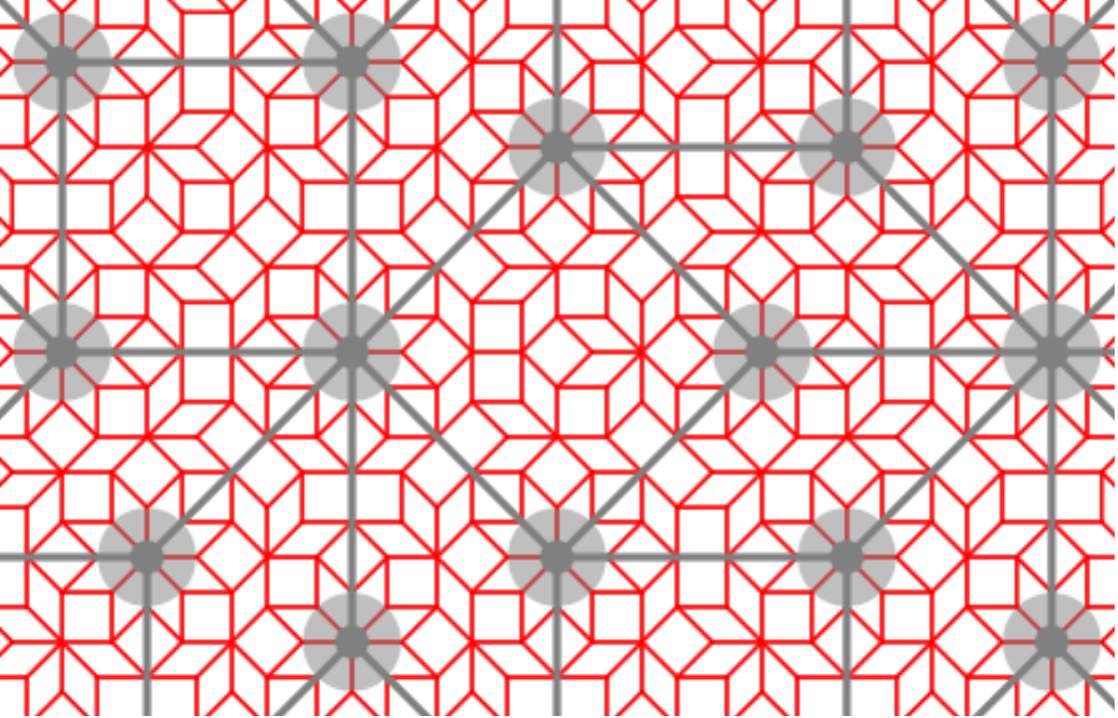} 
    \caption{A portion of the original octagonal tiling in red. Larger dots and grey circles show centers of molecules (model A), and grey bonds show the inflated tiling. }
    \label{fig:modelAmolecules}
\end{figure}

Consider Hamiltonians parametrized by the variable $0\leq \varepsilon\leq 1$, as follows
\begin{eqnarray}
\label{eq:ham2hops}
    H_A &=& (1-\varepsilon)~ H_{mol} +  \varepsilon ~ H_1 \nonumber \\
    H_{mol} &=& - t \sum_{ i \in \alpha_2} \sum_{j(i)} (c^\dag_i c_j + h.c.) 
\end{eqnarray}
where $j(i)=1,..,z$ are nearest neighbors of $i$, the central site of the molecules which belongs to the subset $\alpha_2$ of $z=8$ sites. Fig.\ref{fig:modelAmolecules} shows the original tiling (red) the molecule sites (large dots) and the twice-inflated tiling (gray lines) obtained by connecting the centers of molecules. When $\varepsilon \rightarrow 1$, one obtains the uniform hopping Hamiltonian $H_1$ of Eq.\ref{eq:ham1}. 

\begin{figure} 
    \includegraphics[width=1\columnwidth]{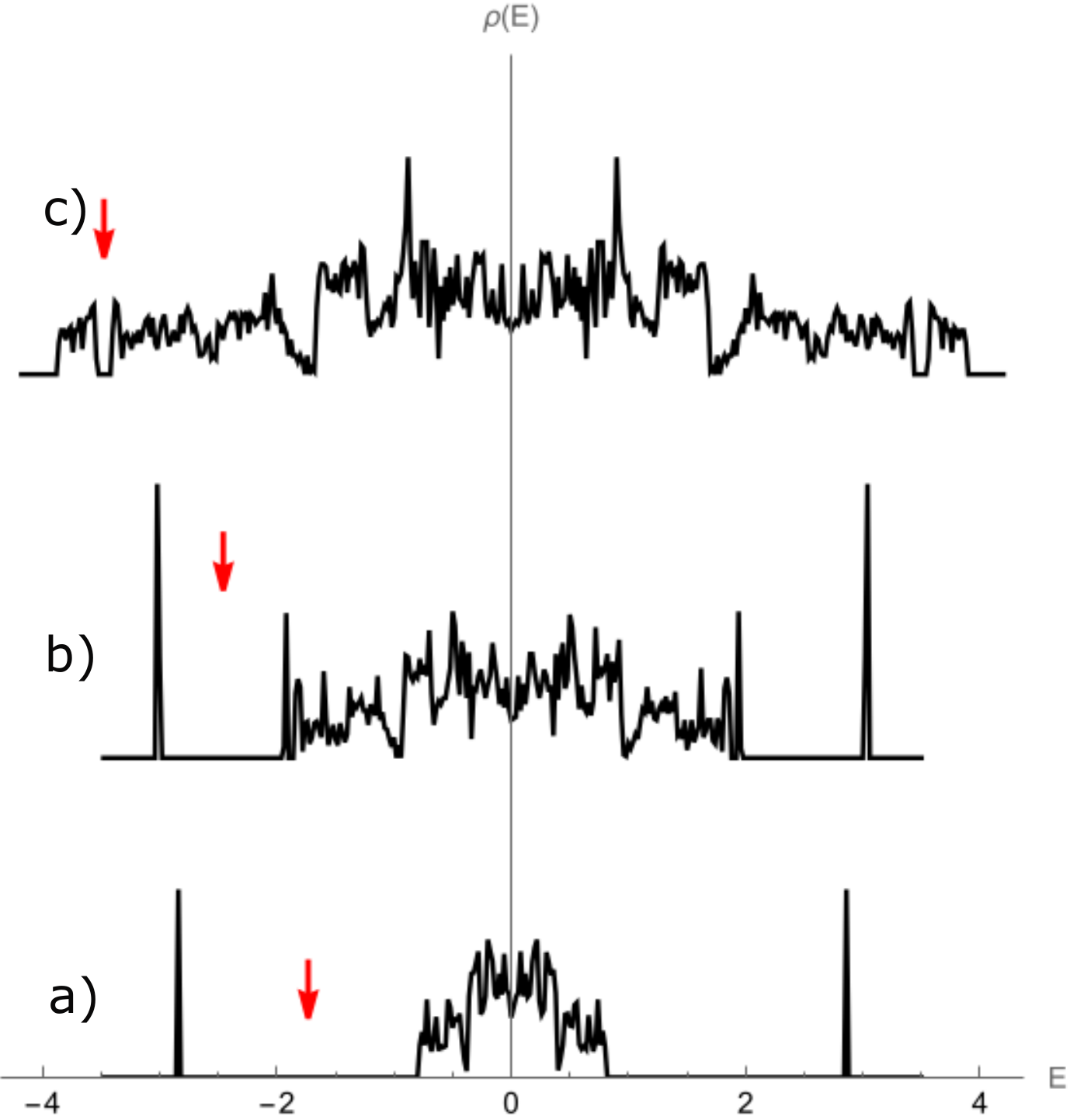} 
    \caption{ Model A DOS as a function of $E$ (in units of $t$) for $\varepsilon=0.2 (a), 0.5$ (b) and 0.9 (c). The DOS have been shifted vertically and rescaled for clarity. The $E=0$ peaks are not shown. The arrows indicate the location of the gap $g_2$ in the lower half of the spectrum. These fractions can be re-expressed in terms of $\sqrt{2}$ in the limit of infinite size.   }
    \label{fig:modelADOS}
\end{figure}

The spectrum for $\varepsilon=0$ consists of three levels: the bonding and antibonding levels $E=\pm 2\sqrt{2}t$, and the central $E=0$ level. The degeneracy of the bonding (anti-bonding) levels is equal to the number of $z=8$ sites. For the $n$th approximant their number is given by $Q_{2n-3}$ (see Table I). For small $\varepsilon$, one can use perturbation theory for degenerate states much as was done for chains in \cite{niunori,kkl}, obtaining three bands. The width of the two lateral bands is given by the number of weak bonds between molecules and is proportional to $\varepsilon^4$. The perturbation theory for the $E=0$ levels (details will be given elsewhere) predicts that the central band is much wider, and $\sim \varepsilon $. Fig.\ref{fig:modelADOS} shows the density of states (DOS) $ \rho(E)=dI/dE$, computed numerically (N=8119 sites) for several values of $\varepsilon$ (plots have been shifted along the vertical axis for clarity). Progressive changes of band widths and the band gaps as $\varepsilon$ is increased can be seen. The arrow shows the location of the principal gap between the bonding band and central band, called $g_2$. Using Table I, one can see that this gap occurs for a band filling of  $Q_{2n-3}/Q_{2n+1}$ (thus a gap index $q=6$), which tends to $\lambda^{-4}$ in the limit $n \rightarrow \infty$. $g_2$ closes linearly as $\varepsilon \rightarrow 1$.  Nb. in this and all subsequent plots, the delta-function at $E=0$ due to localized states is not shown.

\begin{figure}
    \includegraphics[width=1\columnwidth]{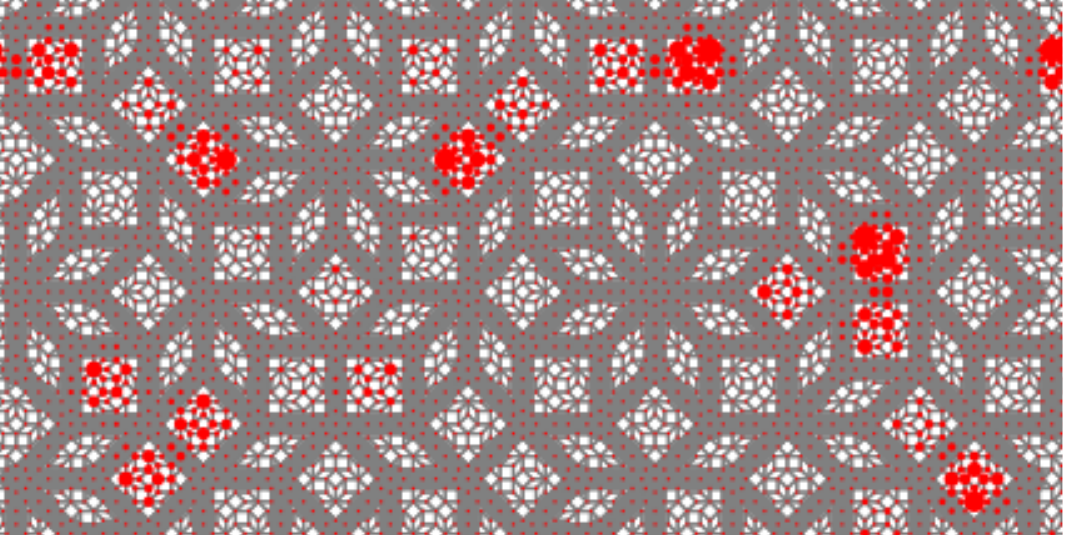}  
    \caption{ Local density of states for $E= -0.315t$. The plot shows the original tiling (thin grey lines), and the inflated tiling obtained on connecting the molecule centers (thick grey lines).Wavefunctions are concentrated within plaquettes of the inflated tiling, because they originate from $E=0$ molecular wavefunctions (see text).}
    \label{fig:modelAwfs}
\end{figure}

In contrast to the wavefunctions of the lateral bands which have largest amplitude on the centers of molecules, the wave functions in the central band are smallest on these sites. Instead they have maximal amplitudes within the plaquettes of the inflated tiling (i.e. on the dual lattice). Fig.\ref{fig:modelAwfs} shows an intensity plot of the local density of states at the energy $E=-0.3615 t$ of the central band (the inflated tiling is shown by thick lines superposed on the original tiling).

\bigskip
\begin{table}
\begin{tabular}{|c||c|c|c|c|c|c|}
\hline 
$z$ & $3$ & $4$ & $5^{(j)}$ & $6$ & $7$ & $8$ \\
\hline 
&&&&&&\\
 $N_z$ & $Q_{2n}$ & $2Q_{2n-1}$ & $Q_{2n-2}$ &  $2Q_{2n-3}$ & $Q_{2n-4}$ & $Q_{2n-3}$ \\
&&&&&&\\
\hline
&&&&&&\\
$f_z$ & $\frac{1}{\lambda}$ & $\frac{2}{\lambda^2}$ & $\frac{1}{\lambda^3}$ & $\frac{2}{\lambda^4}$ & $\frac{1}{\lambda^5}$ & $\frac{1}{\lambda^4}$  \\
&&&&&&\\
\hline
\end{tabular}
\caption{The first row gives the number of sites for a given $z$ as a function of the generation $n$ (see also Appendix A). The second row gives the fraction of each type of site in the infinite tiling. }
\end{table}

\subsubsection{Molecular model B}
\begin{figure}
\includegraphics[width=0.85\columnwidth]{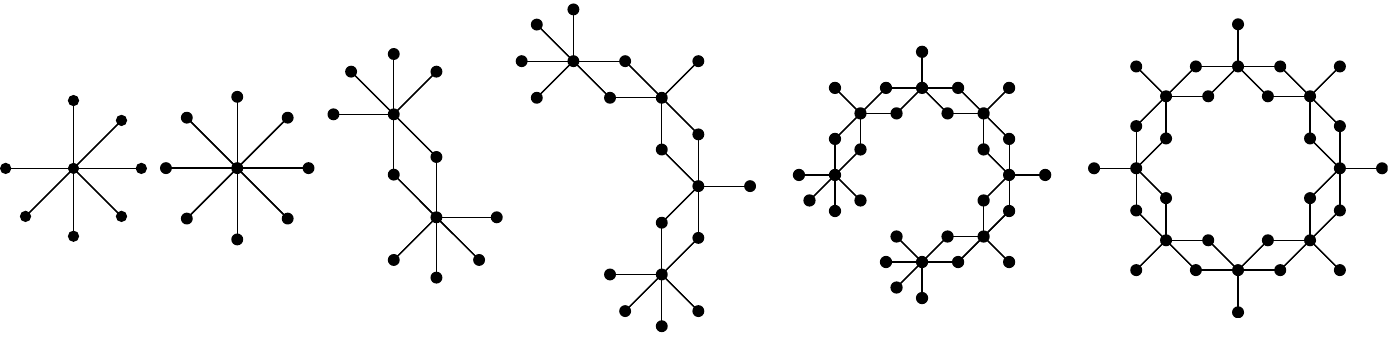}
    \includegraphics[width=0.85\columnwidth]{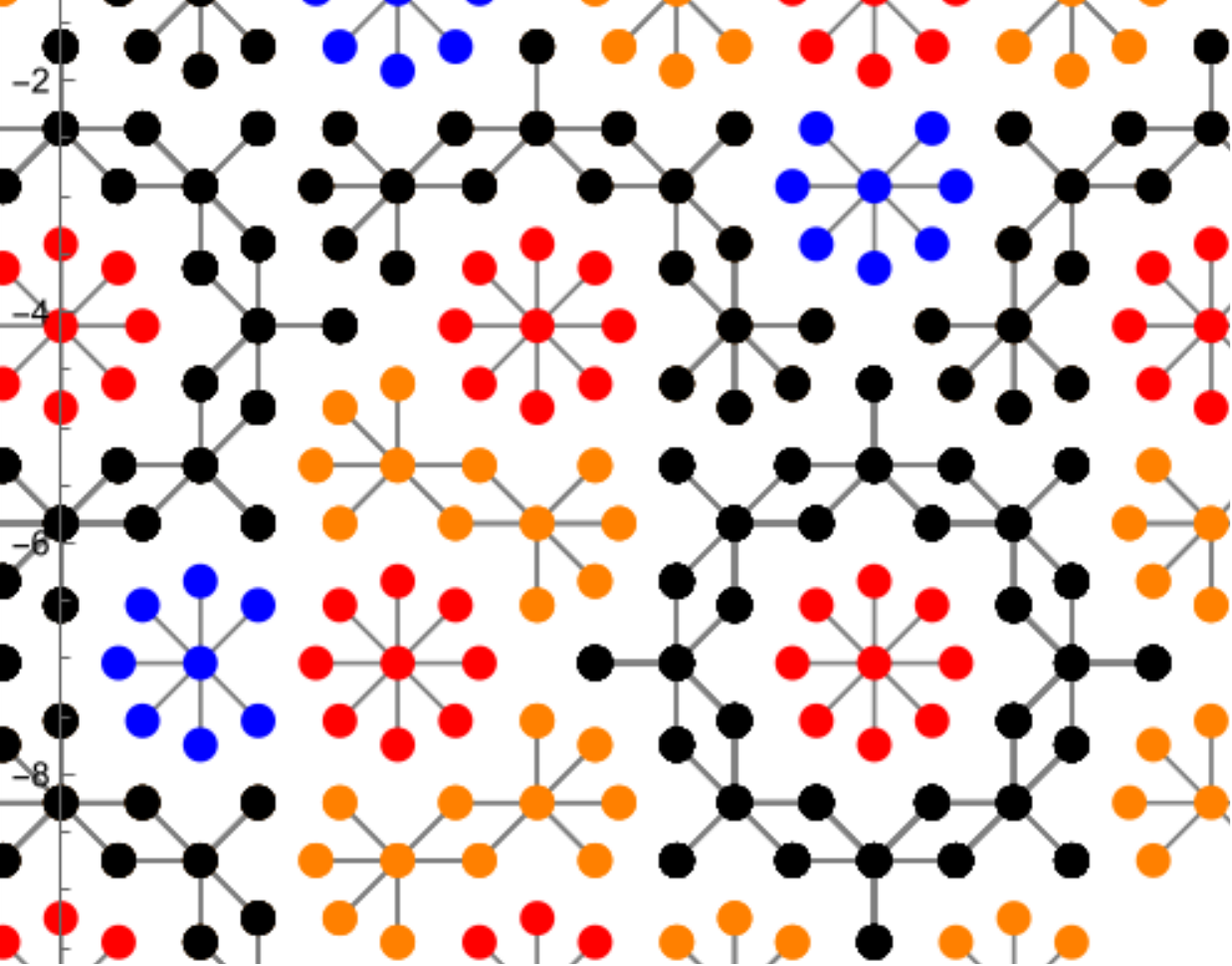}
    \caption{Top: Model B molecules : small molecules (8 or 9 sites) correspond to isolated star-clusters, while larger ones result from clusters that are joined. Bottom:  a portion of the tiling showing how molecules are arranged. The molecules are centered on $z=8$ sites (red), on $z=7$ (blue), $z=6$ (orange) and on $z=5$ and $z=6$ sites (black).  }
    \label{fig:molecule}
\end{figure}

In model B sites of the tiling are first divided into two classes $\alpha$ and $\beta$. The $\alpha$ sites are those which remain after inflation, and become vertices of the inflated tiling and correspond to $5\leq z\leq 8$. The $\beta$ sites are decimated in an inflation, and have $3\leq z\leq 5$. This classification was first used to define block spins in the RG treatment of the antiferromagnetic Heisenberg model \cite{jaga05}. Precise criteria for classification into these two sets and the two types of $z=5$ sites are given in the Appendix. Molecules are clusters of $\alpha$ sites and their nearest neighbors and cover the entire tiling as shown in Fig.\ref{fig:molecule}). The smallest molecules are centered on $z=8$ and 7 (colored red and blue respectively). Larger molecules are formed around sites with $z=6$ and $5$  (colored orange and black respectively).  Note that the set $\alpha_2 \in \alpha$ introduced in the model A are those sites which remain after 2 inflations.  Table 1 (whose derivation is described in \cite{ajduneau}) gives the number of sites of coordination number $z$ in the $n$th approximant. The index $j=1,2$ in $N^{(j)}_5$ refers to two different types of $z=5$ sites. The last line gives the fraction of sites of each type $f_z = N_z/N$ in the infinite size limit. 

\begin{figure}
    \includegraphics[width=1\columnwidth]{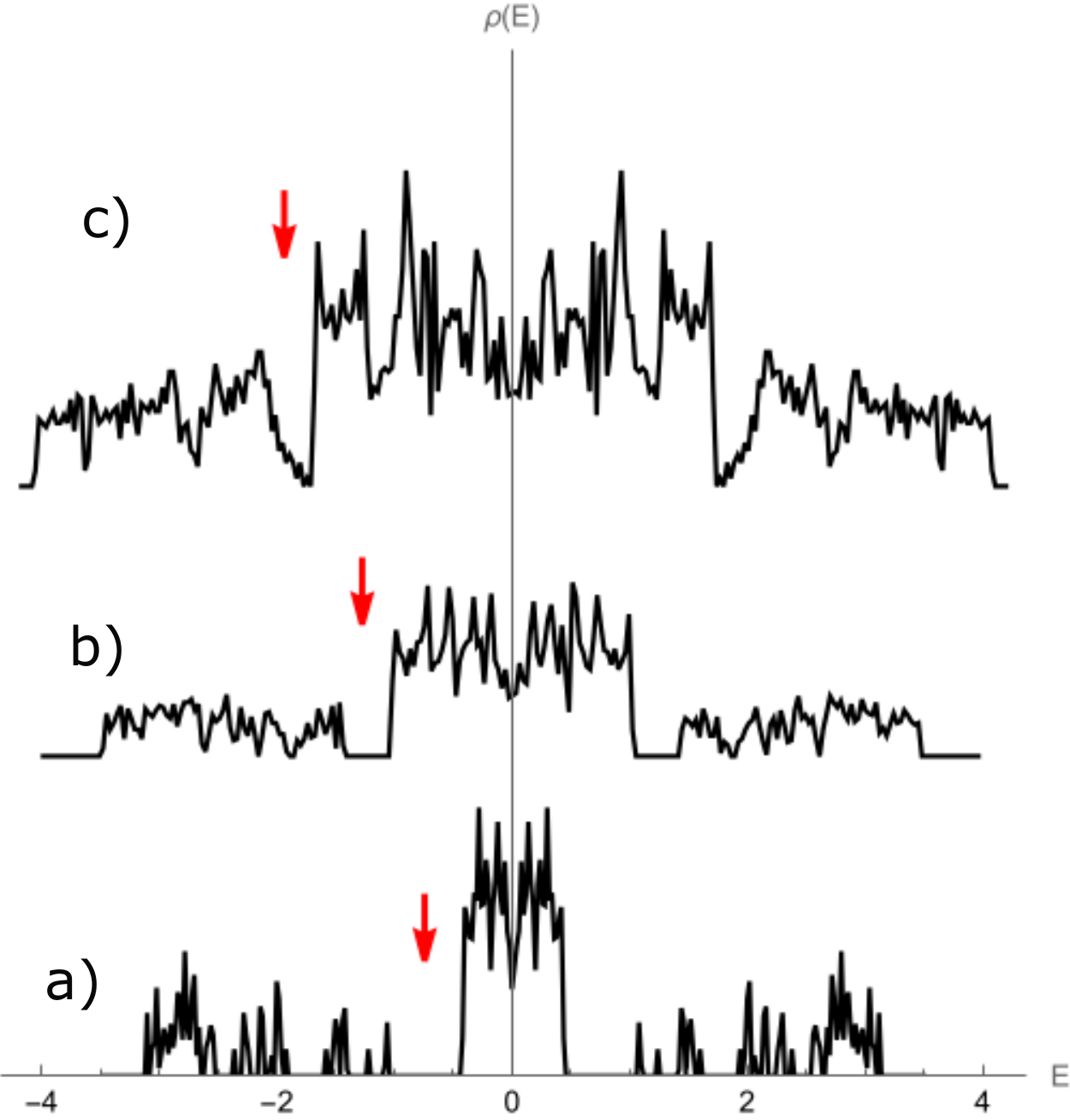} \hskip 1cm 
    \caption{ Model B DOS as a function of $E$ (in units of $t$) for $\varepsilon=0.2$ (a), 0.5 (b) and 0.9 (c). The plots have been shifted vertically and rescaled for clarity. The arrows indicate the location of the gap $g_1$ for the lower half of the spectrum. }
    \label{fig:modelBDOS}
\end{figure}

Consider the family of Hamiltonians given by
\begin{eqnarray}
\label{eq:ham2hops}
    H_B &=& (1-\varepsilon) \tilde{H}_{mol} + \varepsilon H_1 \nonumber \\
    \tilde{H}_{mol} &=& - t \sum_{ i \in \alpha} \sum_{j(i)}{}^{'} (c^\dag_i c_j + h.c.) 
\end{eqnarray}
where the prime in the double sum indicates that sites should not be double counted. For $\varepsilon =0$, this Hamiltonian describes a system of decoupled molecules.  For $\varepsilon =1$, one obtains the Hamiltonian $H_1$ (Eq.\ref{eq:ham1}). As for model A, degenerate perturbation theory gives the bands and the gaps structure for small $\varepsilon$. We illustrate the results obtained by numerical calculations of the DOS in Fig.\ref{fig:modelBDOS} for three values of $\varepsilon$ (system size N=8119). As before, curves have been shifted along the vertical axis for clarity. Narrow bands of discrete bonding and anti-bonding molecular levels for the six different types of molecules can be clearly seen in the bottom curve ($\varepsilon=0.2$). For larger $\varepsilon$ these levels merge progressively, to form a single bonding (anti-bonding) band for $\varepsilon=0.5$. The bonding(anti-bonding) bands are separated from the central band by two main gaps, $g_1$. The corresponding band filling is given by the total fraction of $\alpha$ sites, $Q_{n-2}/Q_n$. Eq.\ref{eq:label} then gives us the values $q=\pm 1$ for the label of gaps $g_1$. This fraction tends to the value $\lambda^{-2}$ as $n \rightarrow\infty$ (see SI).  Indexes of the smaller gaps which occur between molecular bonding states (see for example the lowest curve in Fig.\ref{fig:comparIDOS}) can be similarly determined in terms of the $Q_n$ given in Table 1. Whenever a gap is open, it necessarily corresponds to a filling given by the indexing scheme in Eq.\ref{eq:label}. By continuity, this labeling is expected to hold also in the limit of $\varepsilon=1$.  

For $\varepsilon=1$ the two main gaps close, becoming the main pseudogaps of the hopping Hamiltonian $H_1$ located at the energies $E\approx \pm 1.9t$. The merging of molecular bands explains numerical observations on the structure of the LDOS (local density of states) \cite{jaga94}), and studies of transport, magnetic phases, screening, or superconductivity \cite{passaro,koga,andrade,araujo,takemori}. It explains, for example, why the local charge is largest on high $z$ sites for small filling (occupation of only molecular bonding states) and becomes constant when the Fermi level approaches the band center (filling of center band states). 

\subsubsection{Starting from the atomic limit}
Our third model is described by the Hamiltonian
\begin{align}
\label{eq:modelC}
    H_C =  (1-\varepsilon) H_{at} - \varepsilon H_{1}   \nonumber \\
    H_{at}= t \sum_{i=1}^N z_i c^\dag_i c_i
\end{align}
When $\varepsilon=0$, this Hamiltonian describes $N$ decoupled ``atoms" whose onsite energies depend on their coordination number $E_i=z_i$ \footnote{Even for the decoupled $\epsilon=0$ limit, the onsite potential of each site can be determined from its perpendicular space coordinates, see S.I.}. Particle-hole symmetry is broken. The spectrum is composed of six discrete levels which for nonzero $\varepsilon$ broaden into bands. For $\varepsilon=\frac{1}{2}$, $H_C=H_2$ is the discretized Laplacian model (upto a factor 1/2) that has been studied in \cite{sirebell,benza,zhongmoss}. Further increasing $\epsilon$ leads to the pure hopping model $H_1$ when $\varepsilon \rightarrow 1$. 
Near the atomic limit $\varepsilon \rightarrow 0$, gap positions can be easily determined using Table I. The first large gap --when all $z=3$ sites are occupied-- has IDOS equal to $Q_{2n}/Q_{2n+1}$ which tends to $\lambda^{-1}$ in the infinite tiling, and one can proceed similarly for other gaps. 
    \label{fig:dosmodelC}

\begin{figure}
    \includegraphics[width=1\columnwidth]{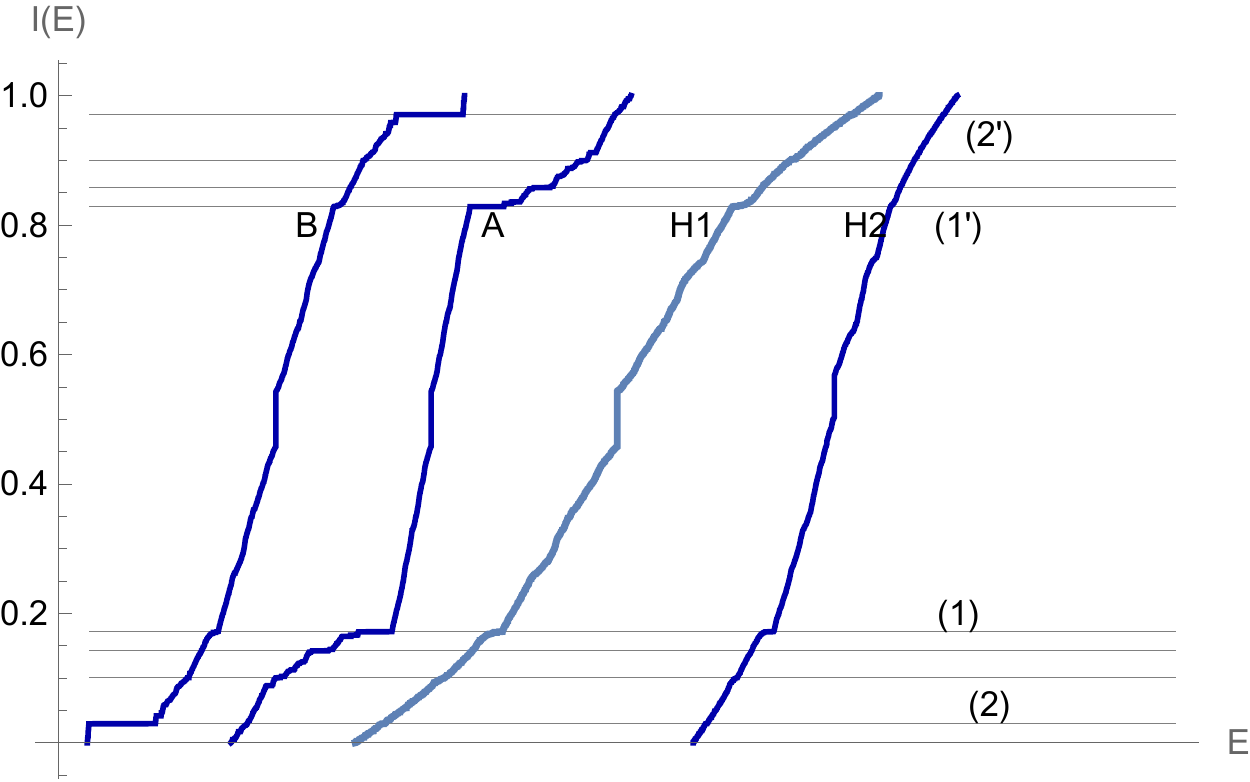} 
    \caption{Comparisons of the IDOS $I(E)$ for: Model A  for $\varepsilon=0.5$, model B for $\varepsilon=0.4$, $H_1$ and $H_2$ (model C for $\varepsilon=0.5$). Horizontal lines show positions of some of the main gaps. (1) and (2) denote the two main pseudogaps (see text). Curves have been shifted along the energy axis for clarity.   }
    \label{fig:comparIDOS}
\end{figure}

Fig.\ref{fig:comparIDOS} shows the topological equivalence of all the above-mentioned models by plotting results obtained numerically for the IDOS for models A, B, C (for the Laplacian $H_2$) and $H_1$. Horizontal lines indicate positions of some gaps present in one or more of the models. The two main pseudogaps of the uniform hopping model are indicated by the labels 1 and 2. All of the models possess localized states which give rise to a step discontinuity of $I(E)$.  

\section{Conclusions} We show how to go from the disconnected molecule or atom limits to the singular continuous quasicrystal limit, and thus gain a better understanding of the electronic properties of the 2D octagonal tiling. Our arguments (details will follow elsewhere) generalize  perturbative 1D analyses introduced for the 1D Fibonacci chain (\cite{niunori,kkl,barache}).   We show a gap labeling scheme for finite systems which is different from the one proposed by Kellendonk and Putnam but does converge to it in the infinite size limit, and is more pertinent for numerical studies or experiments.  

Gap labels were discussed previously for a 2D quasiperiodic system in a magnetic field \cite{jeannoel}. There, the labeling reflects both the algebraic structure of that tiling and winding properties due to the magnetic flux threading the tiles. In contrast, for our present models there is no magnetic flux, and no gaps, only pseudogaps whose locations depend solely on the algebraic structure of the quasiperiodic tiling. Although not a rigorous demonstration, our study plausibly shows how the main pseudogaps of the uniform hopping model arise, when seen in terms of molecular and atomic bands. It explains physical properties, such as the real space distribution of BCS superconducting order parameters \cite{fukushima}. Another example concerns the effects of disorder, when the states near the main pseudogaps localize much faster than the other band states \cite{tarzia}.  

\acknowledgments I thank Pavel Kalugin and Fr\'ed\'eric Pi\'echon for many  discussions and Jean-No\"el Fuchs for helpful comments on this manuscript.

\appendix
\section{Supplementary details on Cut-and-project method for the octagonal tiling and square approximants}
\begin{figure}
    \includegraphics[width=0.75\columnwidth]{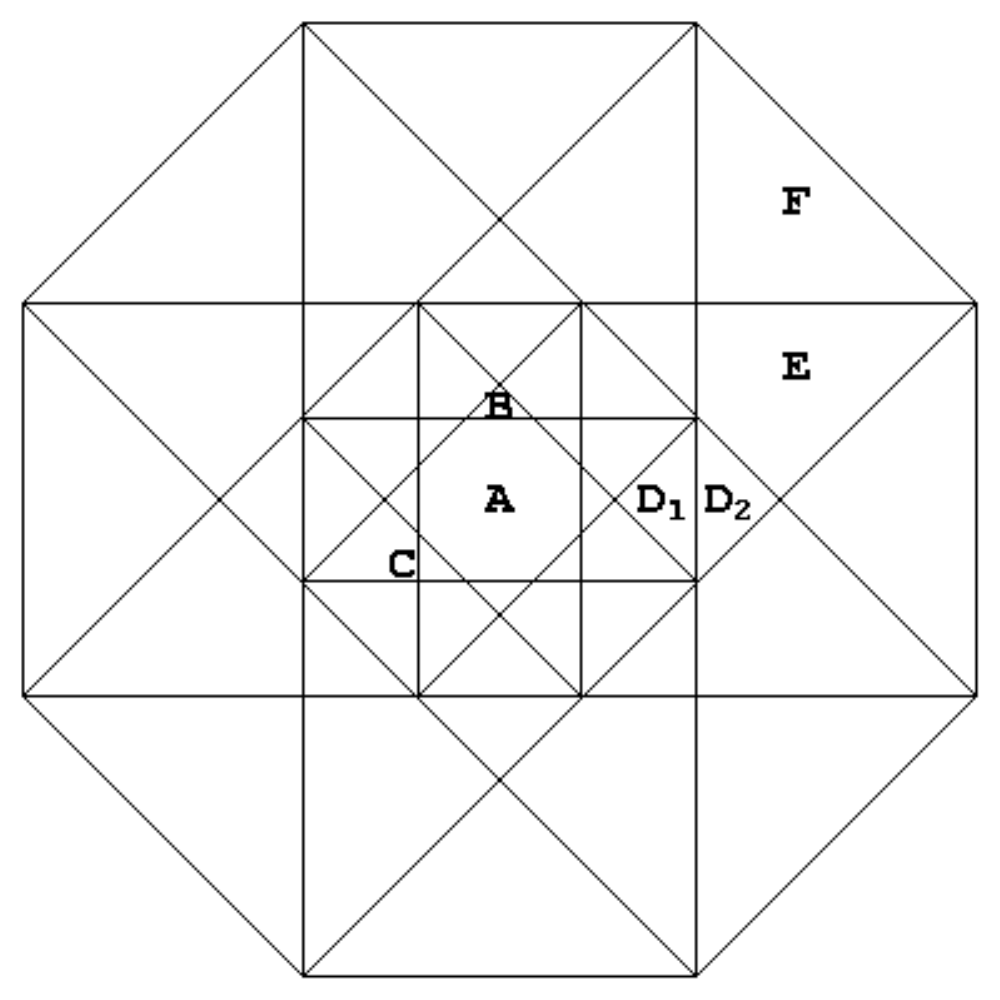}  
    \caption{The octagonal selection window $W$ and its decomposition into subdomains labeled according to the type of site (see text). }
    \label{fig:perpwindows}
\end{figure}

The cut-and-project method \cite{baakegrimm} is a relatively simple way to obtain the infinite tiling and its square approximants \cite{duneau}. We recall briefly the procedure used to generate the infinite quasicrystal first. The parent lattice is a 4D hypercubic lattice $Z_4$ which has two orthogonal sublattices in 2D planes, called the ``physical" and the ``perpendicular" planes both having symmetry under 8-fold rotations. These are aligned irrationally with respect to $Z_4$, with their associated projection matrices $\Pi$ and $\Pi'$. Each of the points of $Z_4$ is tested to see if it satisfies the selection rule, if it does, the point is then projected onto the physical plane. A site is selected if its projection onto the perpendicular plane lies within the ``selection window" $W$, an octagon of side $\lambda$, as shown in Fig.\ref{fig:perpwindows}. 

The perpendicular space window can be divided into polygonal subwindows for each of the types of sites as shown in Fig.\ref{fig:perpwindows}. There are altogether 7 different subdomains corresponding to each of the 6 different values of $8\geq z \geq 3$ (indicated in the figure by the letters A,B,C etc -- only one of the subdomains is labeled for each type, others are related by 8-fold rotations). There are two types of sites of coordination number $z=5$ which we call $D_1$ and $D_2$. The former belongs in the $\alpha$ class, the latter in the $\beta$ class. The lower row of Table 1 gives the fractions of sites of each type in the infinite tiling. These are easily computed in terms of $\lambda$ as they simply given by the area of the corresponding subdomains.

This procedure must be modified in order to obtain approximants, which are periodically repeating structures in the plane \cite{duneau}. The selection window is a distorted octagon, formed by the intersection of two squares. To obtain approximants of different order $n$, the sides of these squares are chosen to be $a(n)=(2\sqrt{2}/\mathcal{N}(n))P_{n+1}$ and $b(n)=(\sqrt{2}/\mathcal{N}(n))(P_n+P_{n+1})$. Here $\mathcal{N}(k)=\sqrt{(4P_n+(-1)^n)}$ is a normalization factor. Table 1 of the paper can be obtained from computing subdomain areas. In the limit $n \rightarrow \infty$, the $a(n)$ and $b(n)$ tend to $\lambda$, and their intersection gives a perfect octagon.
Upon projection one now obtains a pattern which repeats in the 2D plane, with a unit cell of $N^{(n)}$ sites. For the series of square periodic approximants that we have considered, there is a recursive relation giving the number of sites of successive approximants. Let $N^{(n)}$ be the total number of sites and $N^{(n)_3}$ the number of $z=3$ sites in the $n$th square approximant, then
\begin{align}
\left(
\begin{array}{c}
N^{(n+1)}  \\
N^{(n+1)}_3
\end{array} \right)=\left(
\begin{array}{cc}
 5    & 2 \\
  2   & 1
\end{array}\right) \left(
\begin{array}{c}
N^{(n)}  \\
N^{(n)}_3
\end{array} \right)
\end{align}
where the 2 by 2 matrix is the square of the substitution matrix for the Pell numbers. The initial conditions are $N^{(1)}=N^{(1)}_3=1$. Successive approximants thus have the number of sites equal to 1, 7, 41, 239, 1393, 8119,..., The number of $z=3$ sites in successive approximants takes values 1, 3, 17, 99, .... One notes that these numbers are odd and even members respectively of the series $Q_n$.

\begin{figure}
    \includegraphics[width=0.75\columnwidth]{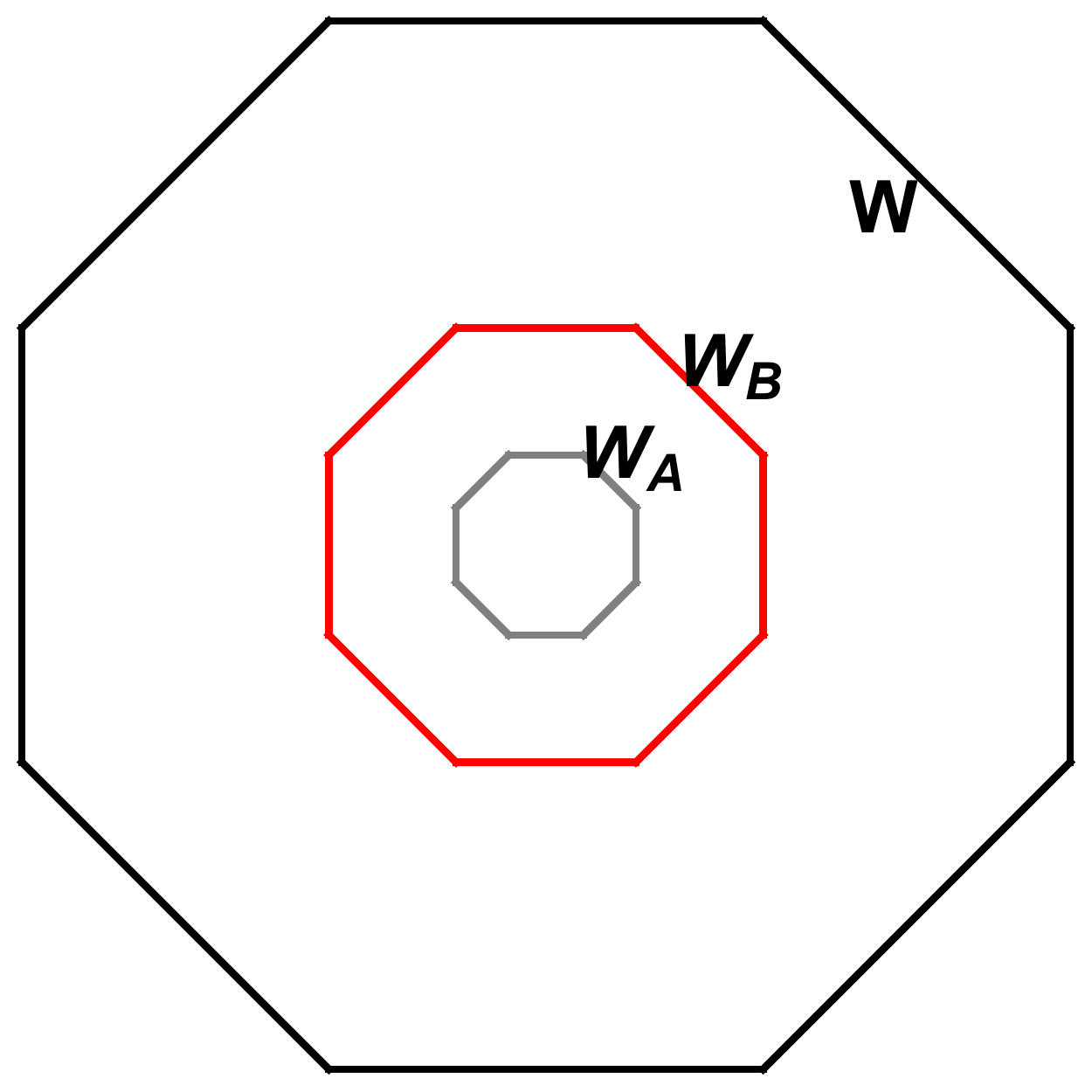} 
    \caption{The selection windows W, $W_A$ and $W_B$ for all sites, $\alpha_2$ sites (model A) and $\alpha$ sites (model B). }
    \label{fig:perpwindowsAB}
\end{figure}

Fig.\ref{fig:perpwindowsAB} shows the selection window $W_A$  corresponding to the ensemble $\alpha_2$ of model A, and the selection window $W_B$ to the corresponding to the ensemble $\alpha$ of model B. They are obtained by shrinking $W$ - for an infinite system the rescaling is given by the factors  $1/\lambda^2$ and $1/\lambda$ respectively. 
For further details see the review by Jagannathan and Duneau \cite{ajduneau}.

\end{document}